\documentstyle[twoside]{adacta}
\begin{document}

\prvastrana=1
\poslednastrana=16

\def\autor{Z. Hradil, R. My\v{s}ka}
\def\nazov{Reconstruction of diagonal elements...}

\headings{1}{16}

\title{   RECONSTRUCTION OF DIAGONAL ELEMENTS OF DENSITY MATRIX
USING MAXIMUM LIKELIHOOD ESTIMATION}

\author{{ Z. Hradil}\footnote{\email{hradil@risc.upol.cz}}}
{ Department of Optics, Palacky University \\ 17.  listopadu 50,
772 07 Olomouc, Czech Republic}
\author{{R. My\v{s}ka}\footnote{\email{myskar@risc.upol.cz}}}
{Joint Laboratory of Optics of Palack\' y University
       \& Phys. Inst. Czech Acad. Sci., \\
       17.~listopadu~50, 772~07 Olomouc, Czech Republic }

\datumy{15 May 1998}{}

\abstract{
The data of the experiment of Schiller et al., Phys. Rev. Lett.
{\bf 77}(1996){ 2933},   are
alternatively evaluated using the maximum likelihood estimation.
The given data are
fitted  better than by the  standard deterministic approach.
Nevertheless, the data are fitted equally well by a whole family
of states. Standard  deterministic predictions
correspond approximately to the envelope of these maximum
likelihood solutions.
}

\section{Introduction}

Quantum state  provides the complete information
about   quantum systems. Recently quantum tomography has been
devised for prediction of quantum state on the basis of
homodyne detection with rotated basis of quadrature operators
[1,2,3].
The technique has been applied to analysis of realistic
measurement and now, quantum state  reconstruction is routinely
used in various applications [4].

Nevertheless, potential problems of
deterministic schemes has been reported.
The positive definiteness of the reconstructed
density matrix is not guaranteed within deterministic data inversion
yielding some nonphysical predictions.
Positive definiteness can be preserved using information theory
[5].
The approach based on the maximum  likelihood (MaxLik) estimation
is closely related to the standard treatment. Instead of the
question: ``What quantum state  is determined by these data?"
the question consistent with quantum theory reads: ``What quantum
state seems to be most likely ?" The general formulation of the
MaxLik  problem  was  given  in [6,7].    The
extremum density matrix is given by nonlinear operator
equation
\begin{equation}
  \hat R(\hat \rho) \hat \rho =  \hat \rho,
\label{vysledek}
\end{equation}
where
\begin{eqnarray}
\hat R = \sum_{i} \frac{f_i}{\rho_{ii}}  \hat \Pi_i,\;\;\;
\label{er}
\rho_{ii}  = {\rm Tr}(\hat \rho \hat \Pi_i)  .
\end{eqnarray}
Here $\hat \Pi_i$ represents  the projectors   corresponding to
(in general unsharp) nonorthogonal measurement.
The measured  relative frequencies are denoted here as $f_i, \sum_i f_i =
1.$
The state may be reconstructed on the subspace where the
projectors provide the resolution of identity operator $\hat R =
\hat 1$.
Although the form of equation (\ref{vysledek}) suggests an
 iterative solution, it is not easy since the
iterations need not be  convergent in general.
Solution simplifies significantly provided that the projectors
$\hat \Pi_i$ commute. The diagonal elements of density matrix
in common commuting basis may be estimated  very effectively,
as showed by Banaszek [8,9]. This numerical approach
will
be used here for evaluation of the diagonal elements of density
matrix for the experiment reported  by Schiller  et.~al [3].
All the necessary steps of general reconstruction scheme
 will be demonstrated on this
example.

\section{Reconstruction}

The diagonal elements of density matrix only
will be reconstructed.
Data corresponding  to the random--phase homodyne detection
[10] are sufficient for this purpose.
Projectors enumerated by position  are given as
\begin{eqnarray}
\hat \Pi(x) =  \frac{1}{2\pi} \int_0^{2\pi} d\theta | x,\theta\rangle
\langle x, \theta |.
\end{eqnarray}
They  commute and are complete on the full interval
$x\in(-\infty,\infty).$
Nevertheless, any realistic measurement will register only a finite
sampling of discrete   decomposition.
Denoting the position of  a particular bin in x--coordinate as
$x_i,$ the projectors in number state basis read
\begin{eqnarray}
\hat \Pi(x_i) =  \Delta x
\sum_{n=0}^{\infty}\sum_{k=0}^{n} \Phi_{k}(x_i) \eta^{k} (1-\eta)^{n-k}
{ {n} \choose {k}}  |n\rangle \langle n|, \\
 \Phi_k(x)= \frac{1}{ 2^k k!\sqrt{\pi} }e^{- x^2} H^2_k(x).
\end{eqnarray}
Here $\eta $ denotes  the efficiency of counting of
photoelectrons, enumerated by index $k$ and $\Delta x $ is the
width of the bin.
This  detection of  discretized
quadrature components    reproduces the
identity operator as
\begin{equation}
\hat R =  \sum_i \hat \Pi(x_i).
\end{equation}
Diagonal elements of $\hat R $ in number state basis
 shows the  subspace where this operator equals approximately
to identity. Here the  reconstruction may be done applying the
MaxLik procedure.  Relative entropy
(log of likelihood function)
\begin{equation}
  K(\rho/f)  =  - \sum_i f_i \ln \frac{\rho_{ii}}{f_i} \ge 0
\label{vysledek3}
\end{equation}
shows how the given state approaches the ideal condition
$\rho_{ii} = f_i.$
Provided that this condition is met, the relative entropy
equals to zero.
In the following the relative entropy will be evaluated
for various states and given  data $f_i.$
For comparison, the relative entropy will  be expressed
in $\% $ with respect to the   entropy of measured data, $S(f) = - \sum_i f_i \ln f_i.$

The data corresponding to  measurement of coherent and squeezed
state will be  considered explicitly.
All the calculations has been done for an overall efficiency
$\eta = 0.85 $ here.
The Fig.~1  plots the  data measured by random--phase  homodyning
 and   reconstructed data using deterministic
and MaxLik approaches.   The subspace for
reconstruction is deliminated  by the dimension about $n_{edge} = 50$
as follows from the  plot of the diagonal elements $R(n)$
in the Fig.~2.
The reconstruction of diagonal elements of density matrix in
number state basis  is plotted in the Fig.~3.
Deterministic  reconstruction yields  slightly nonphysical results
indicated by several  negative diagonal elements obtained.
A typical MaxLik estimation of  diagonal  elements
  is plotted in the upper right panel.
 However, the estimates
depend on the
starting point of the iteration process.
  The procedure has been repeated
$100$ times with
different starting points. The average of all estimates is
plotted in left lower panel of the Fig.~3.
The MaxLik  estimations fits the measured
data obviously better. While the
relative entropy for the deterministic  ``state" is about
 $K(\rho/f) \approx 1.6 \% $ of the value  $ S = 4.717,$
the relative entropy of the  MaxLik estimates  fluctuates around the value
 $ 0.31 \%  $ of the  entropy $ S$ only.
The   histogram  of relative entropies
is plotted in the left upper panel of the Fig.~4.
Significantly, all these states fit well the input data
and there is no observable difference
in fitted data statistics.
Other panels show  the uncertainty of moments
for various MaxLik estimations.
 Moments are normalized with
respect to the  average of all the MaxLik estimates (denoted by
index $~_{AV}$). The
x--coordinate  represents   the deviation of the  moments
$\langle \hat n^k\rangle / \langle \hat n^k\rangle_{AV} -1  $
for $k=1,10,50.$
  All the low moments are estimated  very sharply
within the accuracy $10^{-2}\%.$
However, the $ 50 $--th  moments already  fluctuate within $30 \% .$
The average  number of particles is estimated
here as $ \bar n_{MaxLik} \approx 29.6.$  The value obtained  from the
deterministic approach   is  $ \bar n_{det} \approx 25.6,$ i.e.
$30.1$ considering efficiency $\eta.$

Similar analysis may be done for the data corresponding to
squeezed state. The histogram of random-phase  homodyne detection
and reconstructed data are plotted in the Fig.~5.
The dimension of subspace for
reconstruction is  about $n_{edge} = 100$  (not plotted here).
The reconstructed diagonal elements are  shown in the Fig.~6.
The  deterministic  reconstruction has been considered  on too
small subspace as seen on the upper left panel.
 The MaxLik estimation has been  done  $100$ times on a  $75$
dimensional subspace,          because
 the higher dimension is out of the range of our program.
While deterministic estimation is characterized by the
relative entropy
$K_{sq}(\rho/f) \approx 15 \%$ of the value $ S = 4.02,$
the relative entropy of MaxLik estimates fluctuating around the
value $0.15 \%$ are considerably better.
The histogram of relative entropies  is plotted
in the upper left panel of the Fig.~7, the other panels
show the uncertainty in moments of various   MaxLik estimates.
The value of moments is again related to the
 averaged MaxLik estimate as in the Fig.~4.

\section{Conclusion}

The MaxLik procedure fits the measured data better
than the deterministic scheme, but the results are not single
valued. Instead of a single
state predicted by the deterministic scheme, there is a
 family of states
 fitting  the data equally well.
Hence the MaxLik state reconstruction is more uncertain
in comparison to the deterministic prediction.
However, this uncertainty corresponds to the
probabilistic nature of quantum theory. Measured data do not determine
an unique state and the reconstruction has to take it into
account.

\medskip

\noindent {\bf Acknowledgments}
This work was supported by the
 TMR Network ERB FMRXCT 96-0057 ``Perfect Crystal Neutron Optics''
of the European Union and by the grant of Czech Ministry of Education VS
96028. We are grateful to S. Schiller and G. Breitenbach
 for providing us their data.

\small
\kapitola{References}
\begin{description}
\itemsep0pt

\item{[1]}  \refer{ K. Vogel, H. Risken}{Phys. Rev. A} { 40}
{1989}{ 2847}

\item{[2]} \refer{  D. T. Smithey, M. Beck, M. G. Raymer, A.
Faridani} { Phys. Rev. Lett.} { 70}{1993} {1244}

\item{[3]} \refer{ S. Schiller, G. Breitenbach, S. F. Pereira, T.
M\"{u}ller, J. Mlynek}{ Phys. Rev. Lett.}{ 77}{1996}{ 2933}

\item{[4]}{U. Leonhardt}: {\sl Measuring of the Quantum State of
Light},
{Cambridge Press},{ 1997};

\item{[5]}\refer {V. Bu\v{z}ek, G. Adam, G. Drobn\'{y}}
{ Phys. Rev.  A}{ 54}{1996} {804}

\item{[6]} \refer{ Z. Hradil}{ Phys. Rev.  A}{ 55}{1997} {R1561}

\item{[7]} { Z. Hradil, J. Summhammer, H. Rauch}:
{``\sl Quantum tomography as normalization
of incompatible observations"}, submitted to { Phys. Rev. Lett.}

\item{[8]} { K. Banaszek}:{``\sl Maximum--likelihood estimation
of photon number distribution from homodyne statistics"},
 to appear in {Phys. Rev.  A}{1998};

\item{[9]}{ K. Banaszek},{``\sl Reconstruction of photon
distribution  with positivity constraints"},
to appear in {acta physica slovaca}{1998};

\item{[10]} \refer { M. Munroe, D. Boggavarapu, M. E. Anderson, M. G.
Raymer}{Phys. Rev.  A}{ 52}{1995}{ R924}

\end{description}

\newpage
\begin{figure}
\caption{Fig.~1: Homodyne detection and reconstructed data for
coherent state}
\end{figure}

\begin{figure}
\caption{Fig.~2: Diagonal elements of decomposition of identity}
\end{figure}

\begin{figure}
\caption{Fig.~3: Reconstruction of diagonal elements for coherent
state }
\end{figure}

\begin{figure}
\caption{Fig.~4: Histogram of relative entropies and moments of
particle number  operator for various MaxLik estimates. }
\end{figure}

\begin{figure}
\caption{Fig.~5:  Homodyne detection and reconstructed data for
squeezed  state  }
\end{figure}

\begin{figure}
\caption{Fig.~6: Reconstruction of diagonal elements for squeezed
state}
\end{figure}

\begin{figure}
\caption{Fig.~7: Histogram of relative entropies and moments of
particle number operator for various MaxLik estimates}
\end{figure}

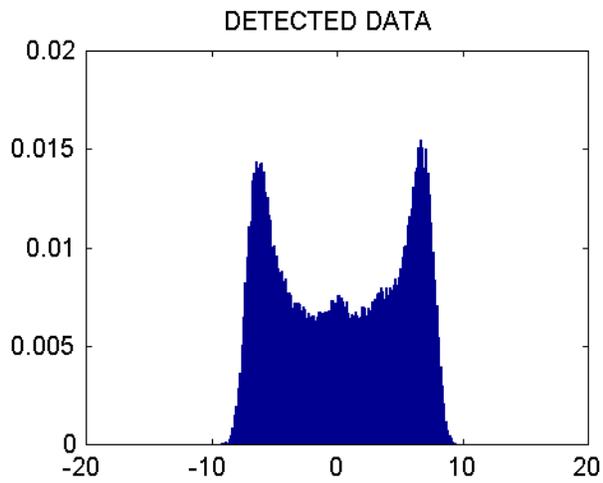
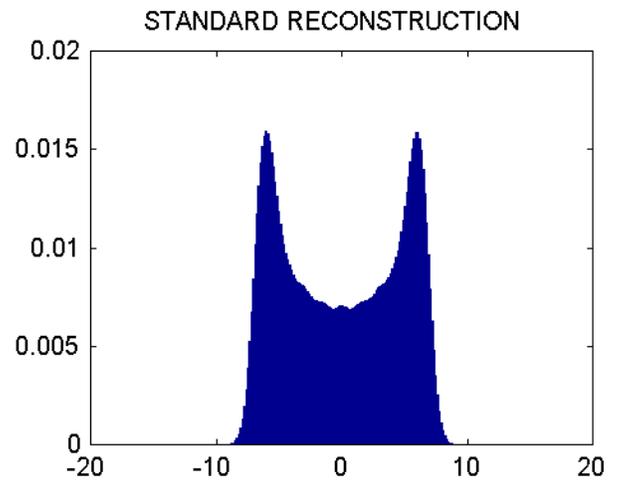
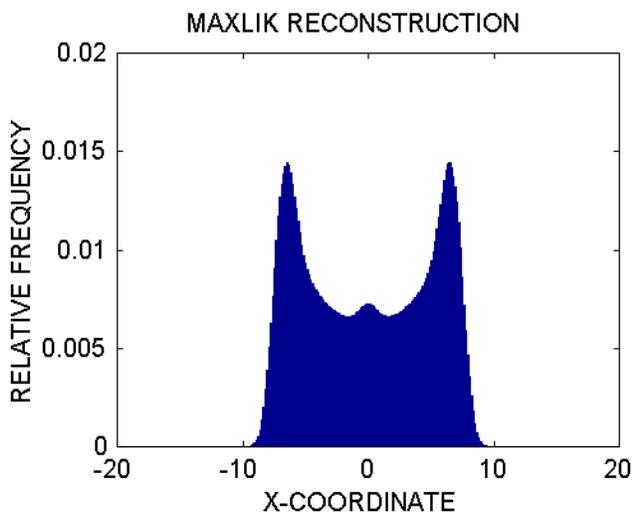
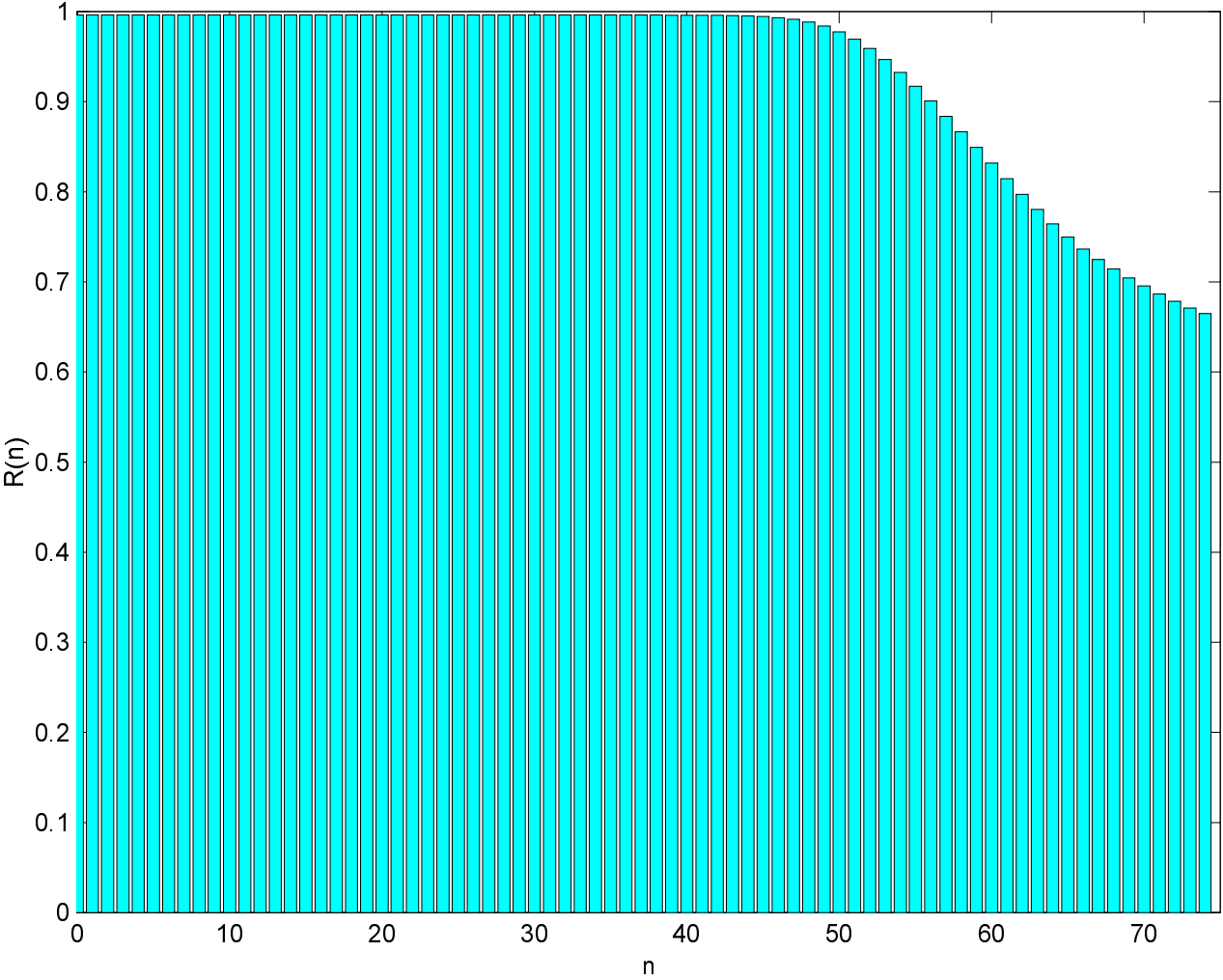
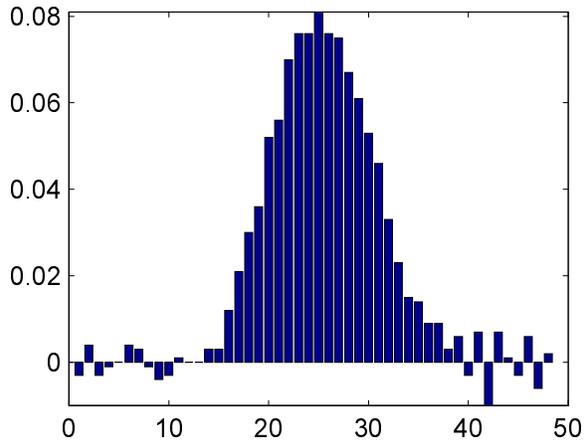
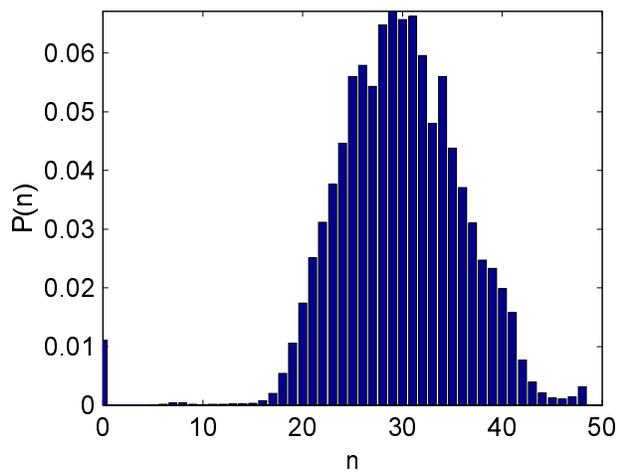
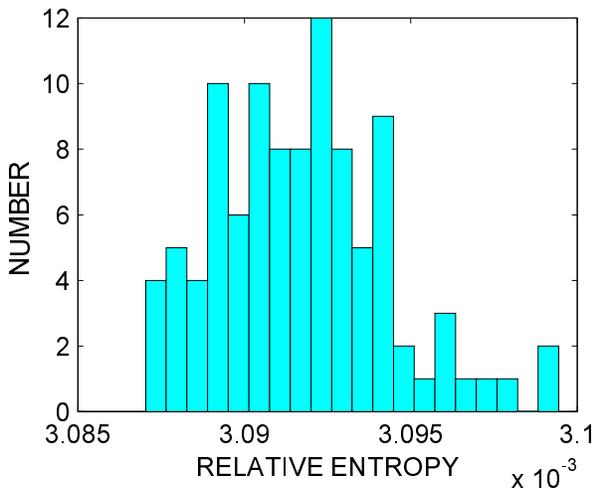
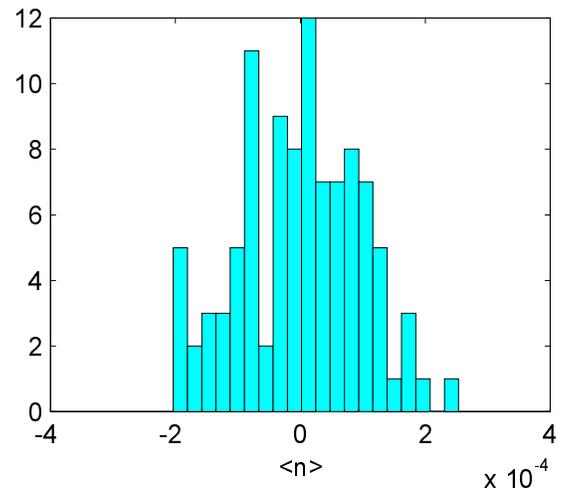
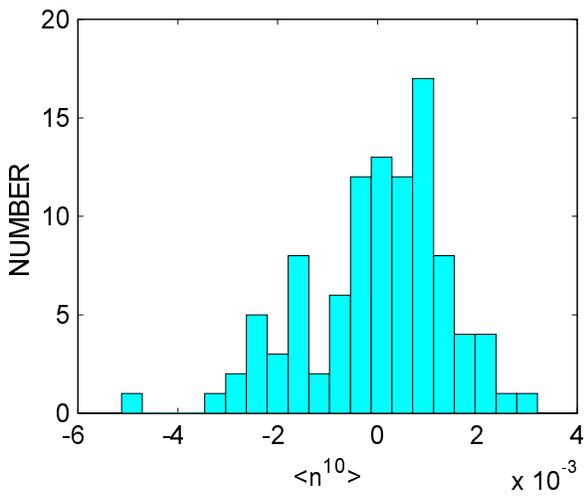
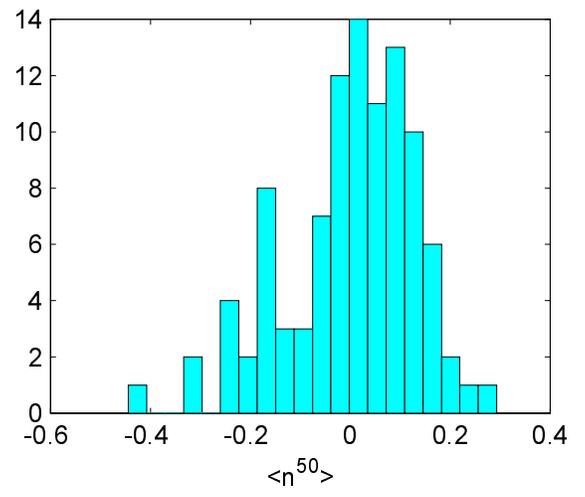
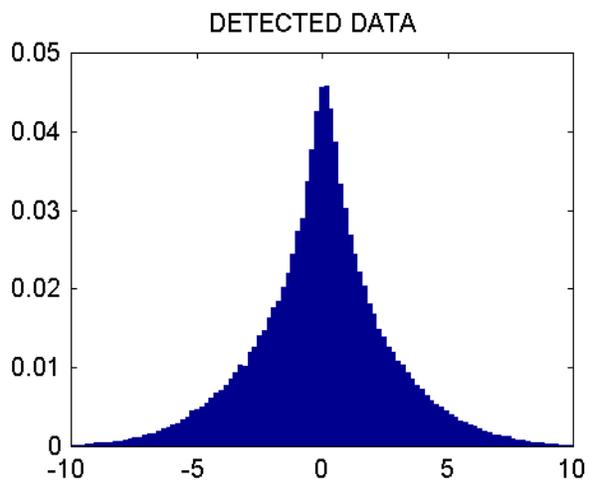
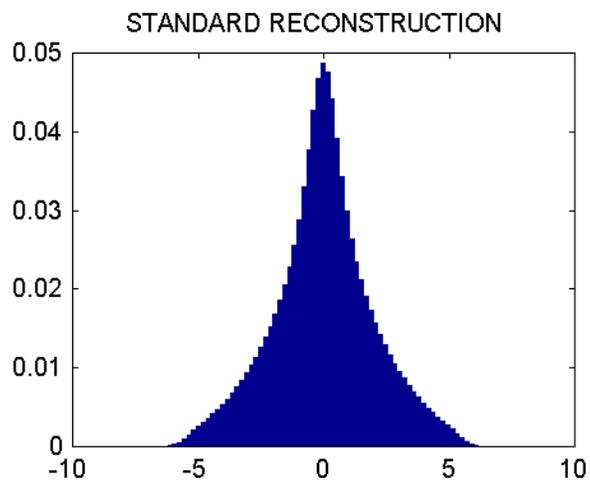
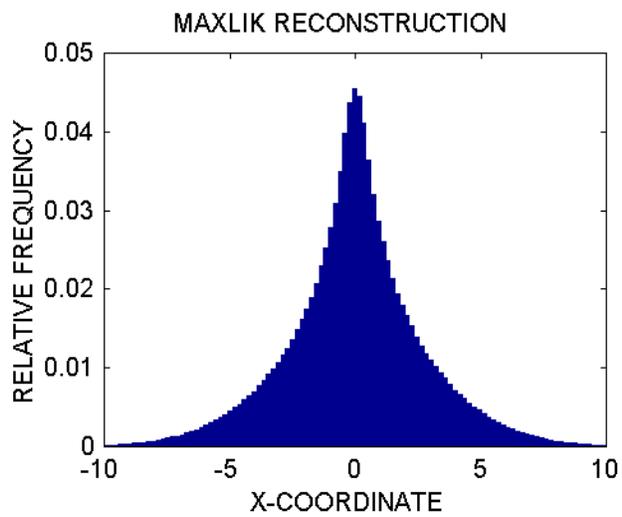
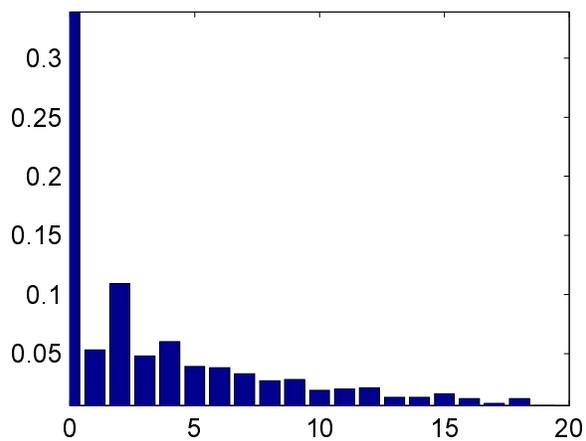
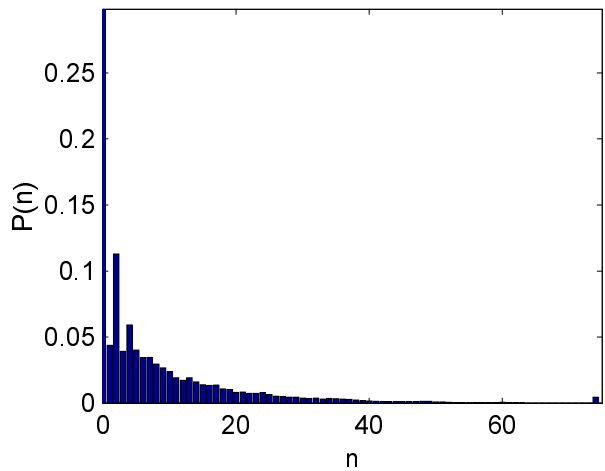
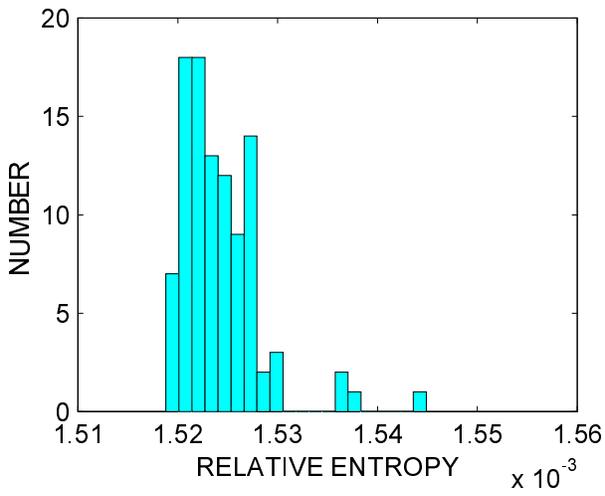
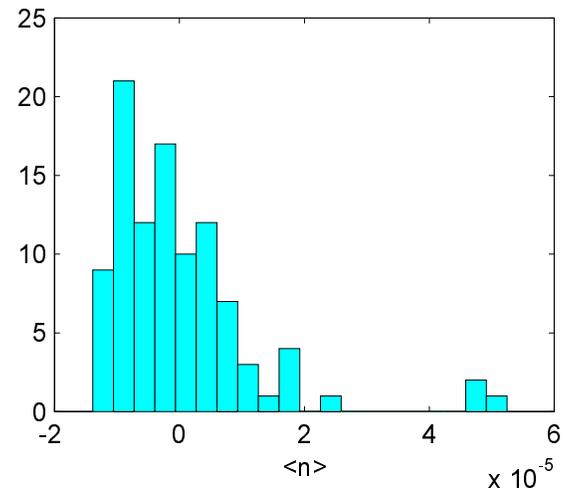
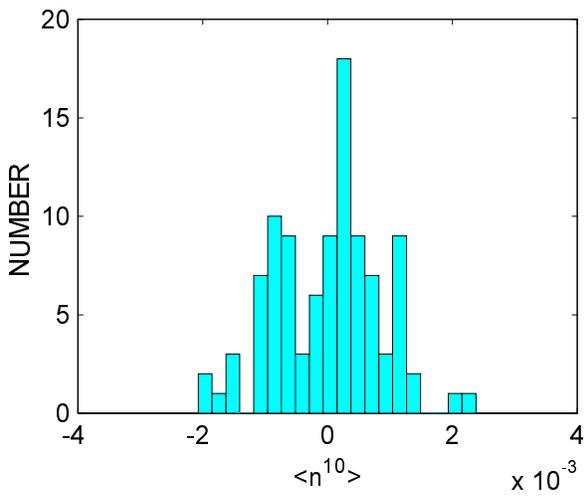
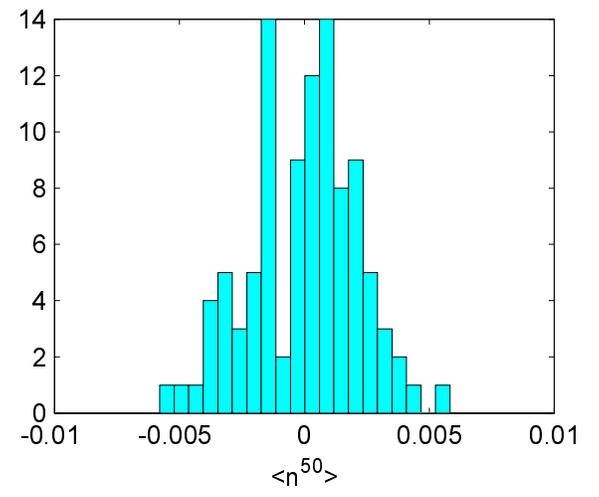
\end{document}